\documentclass[10pt,a4paper]{article}
\usepackage{amsmath,amssymb,graphicx,geometry,cite,indentfirst,textgreek,color}
\begin{document}
\title{Power-law Spectrum of Energetic Particles in Classical Thermal Equilibrium by Pitch-angle Scattering Process}
\author{Yiran Zhang\footnote{Key Laboratory of Planetary Sciences, Purple Mountain Observatory, Chinese Academy of Sciences, Nanjing 210023, China (zhangyr@pmo.ac.cn)}}
\maketitle
\begin{abstract}
The Boltzmann-Gibbs thermodynamic equilibrium state of charged particles pitch-angle scattered by weak plasma waves is discussed. Degrees of freedom of these waves play a fundamental role in constructing the grand canonical ensemble. Via the gyro-resonance condition, fast particles have an inverse break power-law spectrum for $ \varepsilon -\mu \ll T $, where $ \varepsilon $ is the particle energy, $ \mu $ is the chemical potential, $ T $ is the temperature. The break energies are the rest energy and $ -\mu $. For $ \varepsilon \ll -\mu \ll T $, the energy spectral index $ \alpha $ is $ \delta /2+1 $ and $ \delta +1 $ for non- and ultra-relativistic particles, respectively, with $ \delta $ an effective fractal dimension of background magnetic field lines. The spectral index for $ -\mu \ll \varepsilon \ll T $ is $ \alpha +1 $. This thermal equilibrium scenario, combined with the leaky-box model and cosmic-ray observations, seems to suggest that the Galactic magnetic field is super-diffusive with $ \delta \approx 1.4 $.

\end{abstract}
\section{Introduction}
The inverse power-law (PL) spectrum of energetic particles can be found in various physical environments, e.g., in the cosmic space \cite{2006ApJ...640L..79F,2013A&ARv..21...70B,2019SSRv..215...16V} and numerical simulations \cite{2017PhRvL.118e5103Z,2018PhRvL.121y5101C,2020ApJ...893L...7W,2020ApJ...894..136T}. Such a spectrum is conventionally considered as a signal indicating that the system is far from the classical Boltzmann-Gibbs (BG) thermodynamic equilibrium state, with reference generally made to the point of its large deviation from the Maxwell-J\"uttner distribution. Models in the framework of non-thermal processes, mainly based on Fermi acceleration \cite{1949PhRv...75.1169F,1983RPPh...46..973D}, have been proposed to explain the origin of the PL spectrum. The PL distribution can also be produced via generalized thermodynamics with the non-extensive entropy \cite{1988JSP....52..479T}, which implies the breaking of the statistical independence of thermodynamic subsystems (in a particle system) \cite{2008PhRvL.100o5005T}.

As we know, the thermodynamic fluctuation of a subsystem arises from the interaction of a large number of such subsystems \cite{1980stph.book.....L}. In BG statistics, the interaction is so weak that any subsystem is quasi-closed, i.e., the subsystem can completely be described in the absence of the very weak interaction. These subsystems are statistically independent, and the probability of finding any subsystem is $ w\propto e^{-E/T} $, where $ E $ is the subsystem energy, $ T $ is the temperature. It should be noted that, in the field of classical gas dynamics, we are used to treating the gas system as a ``particle system'', i.e., a system whose phase space is defined only with degrees of freedom (DOFs) of particles (i.e., with positions and momenta of particles). If each subsystem is defined as a single particle, i.e., $ E $ represents the particle energy, one has the Maxwell distribution. On the other hand, if the subsystem is defined as a quantum state, i.e., $ E $ is the total energy of a collection of identical particles, one has the Fermi-Dirac or Bose-Einstein (BE) distribution. To avoid any misunderstanding, it should be emphasized that we are not to discuss a quantum gas in this paper.

In the thermal equilibrium state, the short-range interaction of neutral particles can well be interpreted as the particle-particle collision, which is weak enough to maintain the statistical independence, leading to a Maxwell distribution of particles. For a fully ionized plasma, if the frequency of Coulomb collisions is higher than that of the collective electromagnetic field, charged particles can also have a Maxwell distribution \cite{1981phki.book.....L}. These collisions are commonly seen in the low-energy range. As the particle energy increases, both the density and collision cross section decrease, and particles become more likely to interact with the collective field. Since it lacks of an effective relaxation mechanism to produce the Maxwell distribution, one conventionally seeks for solutions from non-thermal kinetics or non-extensive thermodynamics to describe the particle distribution in a ``collisionless'' plasma system.

The above respects can also, however, provide us another viewpoint on thermodynamics of the collisionless plasma. Indeed, the long-range field-particle interaction breaks the statistical independence of particles, it on the other hand implies that DOFs of the field should deeply be involved in the establishment of the equilibrium state. In this sense, only after taking into account both DOFs of the field and particles, one can completely define the quasi-closed subsystem in the framework of classical BG statistics.

Generally, detailed kinematics taking into account the mixture of DOFs of the field and particles may be complicated. Nevertheless, if there is a specific one-to-one relation between the two types of DOFs, the problem may be similar to that of a quantum gas, i.e., the relation can act like a ``wave-particle (WP) duality''. In this paper, we shall consider the case of a pitch-angle (PA) scattering plasma, i.e., use the gyro-resonance (GR) condition to relate the scattering field and particles. Such a condition is actually considered in some literature to couple WP transport equations \cite{2012PhRvL.109f1101B,2018PhRvD..98f3017E}, but it has not been well discussed for a thermodynamic problem. In the following, we shall demonstrate that fast particles can have an inverse break PL spectrum in the classical BG thermal equilibrium state of the PA scattering process.
\section{Kinematic Equation}
It has been shown that charged particles moving much faster than weak Alfv\'en waves (AWs) are primarily subject to a PA scattering process by the waves \cite{1966PhFl....9.2377K,2013A&ARv..21...70B}. On the average, the scattering is most efficient when the first-order GR condition is satisfied, i.e., when the forward traveling gyrating particle co-rotates with the magnetic field perturbation. Since the wave frequency can be neglected compared with the particle motion, in an ensemble of left- and right-hand polarized waves, the primary resonance condition is
\begin{align}
c\left| k_{\parallel}p_{\parallel} \right|=qB,\label{GR}
\end{align}
where $ c $ is the speed of light, $ q $ is the absolute value of the particle charge, $ B $ is the magnitude of the background magnetic field (BMF, i.e., the guide field), $ k_{\parallel} $ and $ p_{\parallel} $ are scalar projections of the wave vector $ \boldsymbol{k} $ and particle momentum $ \boldsymbol{p} $ on the BMF, respectively. Note that one has $ k=\left| k_{\parallel} \right| $ for AWs propagating along the BMF.

Let us assume that the WP interaction exactly occurs at the resonance condition Eq.~(\ref{GR}), which provides a one-to-one map from dynamics of particles to that of wave states. These waves represent characteristic vibrational DOFs of a perturbed magnetic field in a finite (or periodic) volume. For ideal hydrodynamics, it has been shown that a set of phase-space coordinates can be formed in terms of Fourier coefficients of the field \cite{lee1952some}. For convenience, we first consider a steady system in a cubic box with a uniform BMF and unidirectional AWs. The case of bidirectional waves should be given with a linear superposition. The steadiness, in response to equilibrium, implies that the perturbation energy is conserved, which can be used for normalization. Obviously, any Cartesian component $ \psi $ of the perturbation can be expanded into the Fourier series
\begin{align}
\psi =\frac{1}{\sqrt{l^3}}\sum_{s=-\infty}^{\infty}{a_se^{ik_{\parallel s}\left( x-c_{\text{A}}t \right)}},\label{FE}
\end{align}
where $ t $ is the time, $ x $ is the position along the BMF, $ l $ is the side length of the box, $ c_{\text{A}} $ is the Alfv\'en speed, $ a_s $ is the expansion coefficient, $ k_{\parallel s}=2\pi s/l $. Since the integer $ s $ has the implication of marking a wave state, there are $ ldk_{\parallel}/\left( 2\pi \right) $ states in the range of $ dk_{\parallel} $ \cite{1975ctf..book.....L}. Now we assume that the overall system can be divide into small local systems, each of which (approximately) has a uniform BMF. The total number of states in $ dk_{\parallel} $ is then proportional to the total length $ \sum{l} $ of the local systems connected in series on the large-scale global BMF, i.e., proportional to the length along the global field lines, which can generally be tortuous. Denoting the length scale of the overall system measured with a straight stick by $ L $, in a fractal concept one expects the field-line length to obey the scaling law $ \sum{l}\propto L^\delta $ with $ \delta >0 $ the fractal dimension. This law is in fact consistent with anomalous diffusion of chaotic field lines, i.e., $ \delta <2 $ and $ \delta >2 $ represent super- and sub-diffusion, respectively \cite{1995PhPl....2.2653Z,2014PhPl...21g2309R}. In the sense of the dimension, intuitively, the number of states in $ d^{\delta}k $ is $ d\nu =\left[ L/\left( 2\pi \right) \right] ^{\delta}d^{\delta}k $. For waves propagating isotropically, using the $ \left( \delta -1 \right) $-sphere surface area formula, we have the $ k $-space density of states (DOS)
\begin{align}
\frac{d\nu}{dk}=\left( \frac{L}{2\pi} \right) ^{\delta}\frac{2\pi ^{\frac{\delta}{2}}}{\Gamma \left( \frac{\delta}{2} \right)}k^{\delta -1},\label{DOS}
\end{align}
where $ \Gamma $ denotes the Gamma function.

Although the classical thermal equilibrium distribution of energy states does not depend directly on detailed kinematics, we give here, for completeness of our representation, the kinematic equation of the above system. For the moment we shall use the Dirac notation to represent the eigenstate as $ |s\rangle $ for convenience. Rewriting Eq.~(\ref{FE}) to $ |\psi \rangle =\sum_s{a_s|s\rangle} $, if $ c_{\text{A}} $ is a constant, obviously we have
\begin{align}
i\frac{\partial}{\partial t}|\psi \rangle =c_{\text{A}}\hat{k}_{\parallel}|\psi \rangle ,\label{SE}
\end{align}
where the operator $ \hat{k}_{\parallel}=-i\partial /\partial x $ satisfies the eigenvalue equation $ \hat{k}_{\parallel}|s\rangle =k_{\parallel s}|s\rangle $. This equation of motion describes waves propagating forward with the phase speed $ c_{\text{A}} $, and is a branch of the classical wave equation $ \partial ^2|\psi \rangle /\partial t^2=-c_{\text{A}}^2\hat{k}_{\parallel}^2|\psi \rangle $. The time reversal $ t\rightarrow -t $ just transforms Eq.~(\ref{SE}) to the other branch $ i\partial |\psi \rangle /\partial t=-c_{\text{A}}\hat{k}_{\parallel}|\psi \rangle $, which describes backward waves. Thus there is no violation of the time reversal symmetry under $ \hat{k}_{\parallel}\left( -t\right) =-\hat{k}_{\parallel}\left( t\right) $, and the state of resonance can completely be described with Eqs.~(\ref{GR}) and (\ref{SE}). This reversibility may be important for the later invocation of the principle of detailed balance.
\section{Statistical Description}\label{SD}
For clarity, let us give an elementary description of statistics specific to the PA scattering problem. In the kinematic system, the normalization condition $ \langle \psi |\psi \rangle =\sum_s{\left| a_s\right| ^2}=1 $ suggests that $ \left| a_s\right| ^2 $ is the probability of finding the pure state $ |s\rangle $. In the concept of thermodynamics, a statistical ensemble is constructed after one changes the system from the above complete description to an incomplete one, i.e., after putting thermal fluctuations into the system. For a thermodynamic system composed of discrete states, the probability distribution can be given by means of the density matrix $ \hat{w}=\overline{|\psi \rangle \langle \psi |} $, where the overline refers to averaging over the fluctuations. The matrix element is $ w_{ss'}=\langle s|\hat{w}|s' \rangle =\overline{a_sa_{s'}^*} $, the probability of finding the mixed state $ s $ is thus the diagonal element $ w_s\equiv w_{ss} $. Using Eq.~(\ref{SE}), we have the Liouville-von Neumann equation
\begin{align}
\frac{\partial \hat{w}}{\partial t}=\overline{\left( \frac{\partial}{\partial t}|\psi \rangle \right) \langle \psi |}+\overline{|\psi \rangle \frac{\partial}{\partial t}\langle \psi |}=ic_{\text{A}}\left[ \hat{w},\hat{k}_{\parallel} \right] ,
\end{align}
where $ \left[ x,y\right] =xy-yx $ is the commutator. Hence if we study particles traveling with high speeds so that wave propagation can be neglected, i.e., $ c_{\text{A}}=0 $, or distinguish subsystems according to different wave states so that $ \hat{w} $ and $ \hat{k}_{\parallel} $ have a set of common eigenfunctions, i.e., $ \left[ \hat{w},\hat{k}_{\parallel} \right] =0 $, then the probability distribution should be conserved, i.e., $ w_s $ is a constant of motion. This is an obvious consequence of equilibrium.

There are other two important constants of motion for a definite GR state. The one is the particle parallel energy $ \varepsilon _{\parallel s} $ corresponding to the resonance momentum $ p_{\parallel s} $, since the value of $ s $ is given. The other one is the number $ n_s $ of particles associated with the resonance state. As we know, in quantum mechanics the particle number is introduced via the canonical commutation relation. It is not clear whether any similar relation can be introduced into the plasma system, but we recognize the fact of the WP interaction. Then, in view of the reversibility of the kinematic equation, we can directly invoke the principle of detailed balance, which claims that, at each moment of thermal equilibrium, the number of particles scattered out of a resonance state is equal to that scattered into, i.e., $ n_s $ is a constant of motion. Consequently, a thermodynamic subsystem can be defined as a thermally fluctuating collection of particles, which are scattered at a resonance state. Note that the wave energy should also be a constant of motion, which is of less importance for the particle collection if we assume that this energy is independent of fluctuations of $ n_s $.

In classical BG statistics, subsystems are quasi-closed and statistically independent. This requires the probability of finding a composite system to be the product of probabilities of finding the subsystems involved, i.e., $ \ln w_s $ is an additive quantity \cite{1980stph.book.....L}. According to our common understanding, the particle energy and number are also additive quantities. Therefore, the additive constant of motion $ \ln w_s $ can be expressed as a linear combination of $ \varepsilon _{\parallel s}n_s $ and $ n_s $, i.e., we have the grand canonical distribution
\begin{align}
w_s=e^{\frac{\varOmega _s+\left( \mu -\varepsilon _{\parallel s} \right) n_s}{T}},\label{CD}
\end{align}
where the extensive property $ \varOmega _s $ is the grand potential, the intensive property $ \mu $ and $ T $ are the chemical potential and temperature, respectively. The wave energy could be contained in $ \varOmega _s $. Note that Eq.~(\ref{CD}) is the statistical probability of finding a resonance state defined by $ s $.

Following the standard procedure \cite{1980stph.book.....L}, we let $ s $ be fixed to obtain the mean ``occupation number'' $ \overline{n}_s $ of particles at a stationary resonance state. In principle, this occupation is unrestricted, and even if there were any restriction the upper limit of the occupation number could not be a small value. It is thus equivalent to the thermal equilibrium problem of bosons. Under fluctuations of $ n_s $, we have
\begin{align}
1&=\sum_{n_s=0}^{\infty}{w_s}=\frac{e^{\frac{\varOmega _s}{T}}}{1-e^{\frac{\mu -\varepsilon _{\parallel s}}{T}}},\\
\overline{n}_s&=\sum_{n_s=0}^{\infty}{n_sw_s}=-\frac{\partial \varOmega _s}{\partial \mu}=\frac{1}{e^{\frac{\varepsilon _{\parallel s} -\mu}{T}}-1},\label{BE}
\end{align}
where $ \mu <\varepsilon_{\text{m}} $ ensures convergence of the geometric series for any $ \varepsilon_{\parallel s}\geqslant \varepsilon_{\text{m}} $. This means that $ \overline{n}_s $ obeys the BE distribution. The Planck distribution, i.e., $ \mu =0 $, can be derived if one further treats the total particle number as a thermodynamic variable, because $ \mu $ is the derivative of a thermodynamic potential (e.g., the internal energy, enthalpy, free energies) with respect to the particle number, and the principle of minimum energy requires the thermodynamic potential to be a minimum at thermal equilibrium.

To understand the BE distribution, one may on the other hand invoke the well-known counting argument \cite{1980stph.book.....L}. In the case of the PA scattering process, at the state $ s $ there are $ n_s/\overline{n}_s $ neighboring waves, which identically scatter a number $ n_s $ of fast particles at the resonance energy $ \varepsilon _{\parallel s} $. Since each wave can scatter any number (smaller than $ n_s $) of particles, there are $ W_s=\left( n_s+n_s/\overline{n}_s-1 \right) !/\left[ n_s!\left( n_s/\overline{n}_s-1 \right) ! \right] $ ways to distribute $ n_s $ identical particles among $ n_s/\overline{n}_s $ identical resonance states. The closed overall system containing all resonance states, i.e., a system taking all possible values of $ s $ into account, can be seen as a microcanonical ensemble, which is of specified values of the state distribution $ n_s/\overline{n}_s $, total particle number $ N=\sum_s{n_s} $, and total particle resonance energy $ U_{\parallel}=\sum_s{\varepsilon _{\parallel s}n_s} $. At thermal equilibrium, the mean occupation number $ \overline{n}_s $ should maximize the total entropy $ S=\sum_s{\ln W_s} $ under constraints in which $ n_s/\overline{n}_s $, $ N $ and $ U_{\parallel} $ are given. Introducing the Lagrange multiplier $ \mu /T $ and $ -1/T $, we have a necessary equilibrium condition $ \partial \left[ S+\left( \mu N-U_{\parallel}\right) /T \right] /\partial \overline{n}_s=0 $, and then arrive at Eq.~(\ref{BE}) for $ n_s\gg 1 $ and constant $ n_s/\overline{n}_s\gg 1 $.
\section{Thermal Equilibrium Distribution}
For brevity, we shall omit the notation $ s $ if possible. The particle distribution can be obtained with transforming the DOS from the $ k $-space Eq.~(\ref{DOS}) to a $ p $-representation through the GR condition Eq.~(\ref{GR}), and multiplying it by the occupation number Eq.~(\ref{BE}). If waves are random in polarization, i.e., all particles at $ \left| p_{\parallel} \right| $ are identically scattered, the particle distribution with respect to $ p_{\parallel} $ is
\begin{align}
F_{\parallel}=\frac{\overline{n}}{2}\frac{d\nu}{dk}\left| \frac{\partial k}{\partial p_{\parallel}} \right|=\left( \frac{qBL}{2\sqrt{\pi}c} \right) ^{\delta}\frac{\left| p_{\parallel} \right|^{-\delta -1}}{\Gamma \left( \frac{\delta}{2} \right) \left( e^{\frac{\varepsilon _{\parallel}-\mu}{T}}-1 \right)}.\label{DF}
\end{align}
This is the number of particles per unit momentum at $ p_{\parallel} $.

The high speed of particles, i.e., $ v\gg c_{\text{A}} $, and randomness of scatterings tend to isotropize the momentum-space distribution $ f $. The domain of definition of $ f\left( p\right) $ in the momentum space should however be cylindrically symmetric since convergence of $ F_{\parallel} $ requires a minimum value $ p_{\text{m}} $ for $ \left| p_{\parallel}\right| $, below which one may consider that the PA scattering mechanism is no longer valid. Put
\begin{align}
F_{\parallel}=\int_0^{\infty}{f2\pi p_{\bot}dp_{\bot}},
\end{align}
where $ p_{\bot} $ is the perpendicular momentum, it is expected that a primitive function of the integrand is
\begin{align}
-F\equiv \left( \frac{qBL}{2\sqrt{\pi}c} \right) ^{\delta}\frac{p^{-\delta -1}}{\Gamma \left( \frac{\delta}{2} \right) \left( 1-e^{\frac{\varepsilon -\mu}{T}} \right)},\label{PF}
\end{align}
where $ \varepsilon =\sqrt{c^2p^2+m^2c^4}-mc^2 $ is the particle kinetic energy with $ m $ the rest mass, and we simply define $ \varepsilon _{\parallel}=\varepsilon |_{p_{\bot}=0} $. Therefore
\begin{align}
f=-\frac{1}{2\pi p_{\bot}}\frac{\partial F}{\partial p_{\bot}}=\frac{F}{2\pi}\left[ \frac{\delta +1}{p^2}+\frac{c^2}{T\left( \varepsilon +mc^2 \right) \left( 1-e^{\frac{\mu -\varepsilon}{T}} \right)} \right] \label{Df}.
\end{align}

\begin{figure}[b]
	\centering
	\includegraphics[width=0.5\columnwidth]{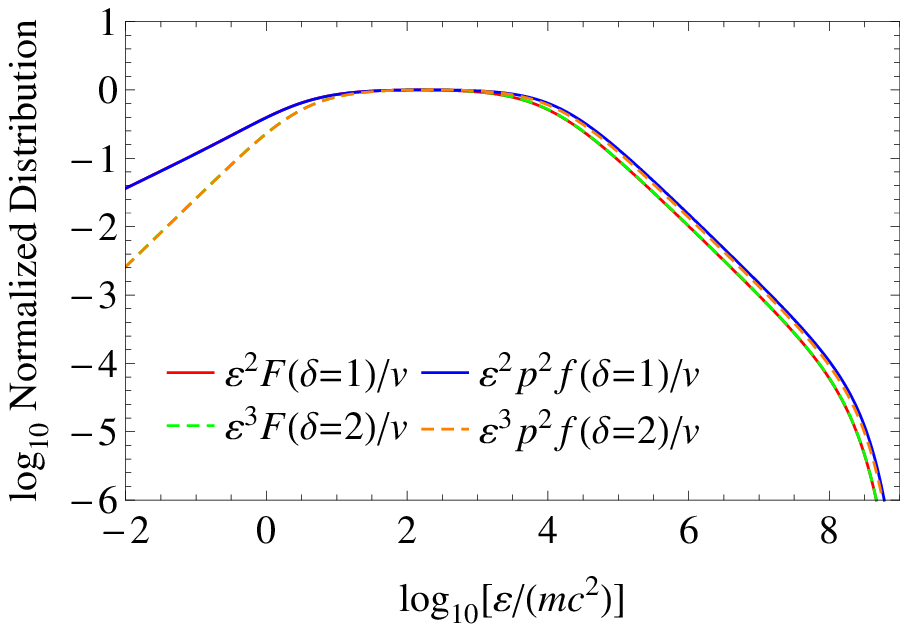}
	\caption{Particle energy distributions obtained with Eqs.~(\ref{PF}) and (\ref{Df}), where $ \mu =-10^4mc^2 $, $ T=10^8mc^2 $, and all the plotted curves are normalized at $ \varepsilon =100mc^2 $. Note that the dependence of $ F $ on $ \varepsilon $ is identical to that of $ F_{\parallel} $ on $ \varepsilon _{\parallel} $.}\label{f1}
\end{figure}

For $ \varepsilon -\mu \ll T $, it is straightforward to find that Eq.~(\ref{Df}) (or (\ref{DF})) reduces to a break PL distribution, i.e., $ f\propto p^{-\delta -3}/\left( \varepsilon -\mu \right) $, which is inversely correlated with $ \varepsilon $ due to $ \delta >0 $. The break energies are $ mc^2 $ and $ -\mu $. We define the spectral index of the energy spectrum as $ -d\ln \left( p^2f/v\right) /d\ln \varepsilon $. Denoting the energy spectral index for $ \varepsilon \ll -\mu \ll T $ by $ \alpha $, then
\begin{align}
\alpha =\left\{ \begin{aligned}
&\frac{\delta}{2}+1,&\text{non-relativistic (NR)}\\
&\delta +1,&\text{ultra-relativistic (UR)}
\end{aligned} \right. .\label{SI}
\end{align}
For $ -\mu \ll \varepsilon \ll T $, the spectral index is $ \alpha +1 $. This spectrum is plotted in Fig.~\ref{f1}.

The physical meaning of the above inverse PL distribution may be understood by analogy to the Rayleigh-Jeans law. The ``inverse'' nature mainly arises from the inverse correlation of $ k $ and $ \varepsilon _{\parallel} $. For $ \varepsilon _{\parallel}-\mu \ll T $, the BE distribution Eq.~(\ref{BE}) reduces to $ \overline{n}=T/\left( \varepsilon _{\parallel}-\mu \right) $, which is a direct reflection of the equipartition phenomenon. As we know, in the equipartition theorem, each DOF of vibration contributes a mean energy equal to the temperature to the system \cite{1980stph.book.....L}. The above subsystems are distinguished exactly by DOFs of waves and each of them is thus of a mean energy $ T $, which is the total effective energy of $ \overline{n} $ particles. In addition, as the chemical potential is the energy required by the system for adding a particle, the particle appears to effectively contribute an energy $ \varepsilon_{\parallel}-\mu $ to the system, i.e., the effective particle energy is $ \varepsilon _{\parallel}-\mu $.

In the opposite limiting case, i.e., $ \varepsilon -\mu \gg T $, the distribution Eq.~(\ref{Df}) (or (\ref{DF})) exponentially cuts off toward high energies. This could be understood by analogy to the Wien approximation, in which subsystems are distinguished effectively by particles, because most microstates are not occupied by a particle, i.e., $ \overline{n}=e^{\left( \mu -\varepsilon _{\parallel}\right) /T}\ll 1 $.

It should be emphasized that the temperature $ T $ is a characteristic energy width of the canonical distribution, i.e., $ T $ should roughly be the mean energy of subsystems, or of a thermally fluctuating subsystem according to the ergodic hypothesis. In our case, $ T $ is the mean effective energy of particle collections, in each of which particles are scattered by waves at a resonance state. The mean energy of particles throughout the overall system depends on the spectral slope, e.g., if $ \varepsilon _{\text{m}}\ll -\mu \ll T $, it is of the order of $ -\mu $ and $ \varepsilon _{\text{m}} $ for $ \alpha <2 $ and $ \alpha >2 $, respectively.
\section{Thermodynamic Quantities}
Extensive properties of the overall system are readily obtained with integration of the distribution function. The total particle number, kinetic energy, parallel energy, (additive) perpendicular energy, entropy are defined as
\begin{align}
N&=2\int_{p_{\text{m}}}^{\infty}{F_{\parallel}dp_{\parallel}},\label{PN}\\
U&=2\int_{p_{\text{m}}}^{\infty}{dp_{\parallel}\int_0^{\infty}{\varepsilon f2\pi p_{\bot}dp_{\bot}}}=2\int_{p_{\text{m}}}^{\infty}{dp_{\parallel}\int_0^{F_{\parallel}}{\varepsilon dF}}=U_{\parallel}+U_{\bot},\\
U_{\parallel}&=2\int_{p_{\text{m}}}^{\infty}{\varepsilon _{\parallel}F_{\parallel}dp_{\parallel}},\\
U_{\bot}&=2\int_{p_{\text{m}}}^{\infty}{dp_{\parallel}\int_{\varepsilon _{\parallel}}^{\infty}{Fd\varepsilon}},\\
S&=2\int_{p_{\text{m}}}^{\infty}{-\overline{\ln w}\frac{F_{\parallel}}{\overline{n}}dp_{\parallel}}=\frac{1}{T}\left( U_{\parallel}-\mu N-2\int_{p_{\text{m}}}^{\infty}{\varOmega \frac{F_{\parallel}}{\overline{n}}dp_{\parallel}} \right) ,
\end{align}
respectively. As we know, in a usual Bose gas the total number of microstates is infinite due to the unbounded $ k $-space volume of the system. In the PA scattering gas, according to the GR condition Eq.~(\ref{GR}), the existence of the minimum momentum $ p_{\text{m}} $ provides a maximum value of $ k $, which imposes a finite boundary on the $ k $-space volume. We can then derive a finite total number of resonance states,
\begin{align}
\nu =2\int_{p_{\text{m}}}^{\infty}{\frac{F_{\parallel}}{\overline{n}}dp_{\parallel}}=\frac{2^{1-\delta}}{\delta \Gamma \left( \frac{\delta}{2} \right)}\left( \frac{qBL}{\sqrt{\pi}cp_{\text{m}}} \right) ^{\delta}.\label{SN}
\end{align}

Obviously, the above extensive properties are all proportional to the system $ \delta $-dimensional volume $ L^\delta $, with which the DOS is measured, showing that the system is homogeneous. Any intensive property of such a homogeneous system should be independent of $ L^\delta $. Especially, the intensive pressure $ \varPi $, i.e., the conjugate of $ L^\delta $, should be defined as
\begin{align}
\varPi =-\frac{2}{L^{\delta}}\int_{p_{\text{m}}}^{\infty}{\varOmega \frac{F_{\parallel}}{\overline{n}}dp_{\parallel}}=\frac{TS+\mu N-U_{\parallel}}{L^{\delta}}.
\end{align}
This definition just means that $ -\varPi L^{\delta} $ and $ \mu N $ are the total grand potential and Gibbs free energy, respectively, provided that the first law of thermodynamics $ dU_{\parallel}=TdS-\varPi d\left( L^{\delta} \right) +\mu dN $ is satisfied. On the other hand, an ordinary gas pressure, e.g.,
\begin{align}
P=\frac{2}{V}\int_{p_{\text{m}}}^{\infty}{dp_{\parallel}\int_0^{\infty}{\frac{vp}{3}f2\pi p_{\bot}dp_{\bot}}},\label{GP}
\end{align}
is generally not an intensive property except for $ \delta =3 $ and the system spatial volume $ V=L^3 $.

We are more interested in the thermodynamic limit $ \varepsilon _{\text{m}} \ll T $, in which the system contains a large number of subsystems, and the particle distribution has a prominent PL shape, implying that most of the subsystems are in equipartition with $ \left( \varepsilon _{\parallel}-\mu \right) \overline{n}=T $. To simplify the equation of state (EOS), let us further consider purely NR or UR particles, and that $ \delta $ is not too small. These allow us to evaluate the above $ N $ and $ U_{\parallel} $ with approximating the integrands as single PL functions. The equipartition now has a macroscopic meaning in the form of $ U_{\parallel}-\mu N=\nu T $. For $ \varepsilon _{\text{m}}\ll -\mu \ll T $, the equipartition reduces to $ -\mu N=\nu T $, after some elementary calculations we have
\begin{align}
U_{\parallel}&=\frac{\alpha -1}{\alpha -2}N\varepsilon _{\text{m}}=\left\{ \begin{aligned}
&\left( \alpha -\frac{3}{2} \right) U_{\bot},&\text{NR}\\
&\left( \alpha -1 \right) U_{\bot},&\text{UR}
\end{aligned} \right. ,\label{UP}
\end{align}
which is valid for $ \alpha >2 $. For $ -\mu \ll \varepsilon _{\text{m}}\ll T $, the equipartition reduces to $ U_{\parallel}=\nu T $, and other EOSs can be derived with $ \alpha \rightarrow \alpha +1 $ in Eq.~(\ref{UP}).

Similarly, taking $ \varOmega =T\ln \left( -\mu /T\right) $ for $ \varepsilon _{\text{m}}\ll -\mu \ll T $, and $ \varOmega =T\ln \left( \varepsilon _{\parallel}/T\right) $ for $ -\mu \ll \varepsilon _{\text{m}}\ll T $, we evaluate
\begin{align}
S=\left\{ \begin{aligned}
&\frac{-\mu N}{T}\left( 1+\ln \frac{T}{-\mu} \right) ,&\varepsilon _{\text{m}}\ll -\mu \ll T\\
&\frac{U_{\parallel}}{T}\left( \frac{\alpha -2}{\alpha -1}+\ln \frac{T}{\varepsilon _{\text{m}}} \right) ,&-\mu \ll \varepsilon _{\text{m}}\ll T
\end{aligned} \right. ,
\end{align}
where the second equation is valid for $ \alpha >1+1/\ln \left( T/\varepsilon _{\text{m}}\right) $ to ensure $ \varPi >0 $.
\section{Discussion}
An appropriate definition of the subsystem is of central importance for a correct description of thermodynamics. Despite the conventional single-particle definition, in consideration of the effect of DOFs of a perturbed magnetic field on the establishment of thermal equilibrium, in this paper we attempt to define the quasi-closed subsystem of a collisionless plasma as a thermally fluctuating collection of fast charged particles, which are of the same resonance energy, i.e., are PA scattered by monochromatic AWs. The most important consequence is that, in the BG thermal equilibrium state without reference to any detailed relaxation process, low-energy particles have an inverse break PL spectrum, which signals that the subsystems are distinguished by waves. The PL spectral index is determined by the fractal dimension $ \delta $, i.e., it depends on the space-filling capacity---in other words, diffusion property---of effective field lines of the global BMF (which is a guide field).

Directly applying the spectral index Eq.~(\ref{SI}) to observational spectra of cosmic energetic particles yields interesting results. It is found that solar-wind NR nuclei with energies typically from keV to MeV have a PL spectrum with the energy spectral index about 2 \cite{2006ApJ...640L..79F}. UR particles in the interstellar medium (ISM), i.e., Galactic cosmic rays (CRs), also obey a PL spectrum, which extends from GeV to tens of EeV with the spectral index close to 3 \cite{2013A&ARv..21...70B}. The direct application of Eq.~(\ref{SI}) to these observations leads to
\begin{align}
\delta \sim \left\{ \begin{aligned}
&\begin{aligned}
&2,&\text{NR \& UR},
\end{aligned}&\varepsilon \ll -\mu \ll T\\
&\begin{aligned}
&0,&\text{NR}\\
&1,&\text{UR}
\end{aligned} \bigg\} ,&-\mu \ll \varepsilon \ll T
\end{aligned} \right. .
\end{align}
This implies that, if the negative of the chemical potential $ \mu $ is small, field lines of the BMF for the CRs are quasi-straight, i.e., $ \delta \sim 1 $, while those for the NR nuclei are likely to be mostly confined in some point-like sources, i.e., $ \delta \sim 0 $. The energy-dependent $ \delta $ may arise from the case that particles with different gyro-radii see different structures of a multi-fractal BMF. However, according to Eq.~(\ref{SN}), as $ \delta \rightarrow 0 $ the total number of subsystems $ \nu \rightarrow 1 $, which may be a violation of the thermodynamic limit. On the other hand, if $ -\mu $ is large, the field lines for the CRs and NR nuclei can be unified to have a quasi-Brownian shape, i.e., $ \delta \sim 2 $.

However, a correction to the above estimates of $ \delta $ may be necessary in consideration of the observed energy-dependent flux ratio of secondary to primary CRs, which implies leakage of diffusive particles from Galaxy in a typical propagation scenario. To be consistent with this framework, we treat the in-situ creation of thermal particles by the PA scattering process as an injection of primary CRs throughout the Galactic CR halo, and assume a finite escape boundary of particles from the system. Integration of the diffusion equation, or simply the leaky-box analysis, shows that the equilibrium spectral index under the injection-escape balance, compared with the injection spectral index, increases by a value reflecting the energy dependence of the diffusion coefficient. Since secondary CRs are created directly from collisions of primary CRs with the ISM rather than from thermal equilibrium, the secondary-to-primary ratio can be explained. Considering the primary CR spectral index to be 2.7, and that CRs obey Kolmogorov diffusion as inferred from the observed sub-TeV boron-to-carbon ratio \cite{2016PhRvL.117w1102A}, we estimate $ \delta \approx 0.4 $ and 1.4 for small and large $ -\mu $, respectively. By definition, these correspond to super-diffusive BMFs, while the former implies that there are non-scattering regions in the system, and the latter, which seems more plausible for CR propagation, gives an intermediate field between the fully uniform and diffusive shape, e.g., the one characterized with a L\'evy flight \cite{1995PhPl....2.2653Z} or Arnold-Beltrami-Childress configuration \cite{2014PhPl...21g2309R}.

The PA scattering scenario of thermal equilibrium may be a candidate for explaining the PL spectrum of galaxy-cluster diffuse radio emission. It is recognized that synchrotron emitting electrons (in GeV energies) that give rise to the radio halo should be produced in-situ, as their typical diffusion length (10 pc) in the characteristic lifetime (100 Myr) is much smaller than the length scale (Mpc) of the halo \cite{2019SSRv..215...16V}. Most observations of such halos predict electron spectral indices in the range of 3--4, corresponding directly to $ 1<\delta <2 $ and $ 2<\delta <3 $ for small and large $ -\mu $, which give a super- and sub-diffusive BMF, respectively. These estimates of $ \delta $ may be compared in further discussion with spectral indices of magnetic turbulence revealed in Faraday rotation studies.

For a given value of $ \delta $, the temperature $ T $ can be calculated with any segment of the PL spectrum, because all subsystems corresponding to the PL spectrum are in equipartition with a mean energy of $ T $. For Galactic CRs (above $ \text{GeV}\gg \varepsilon _{\text{m}} $), the particle number $ N\sim 0.5U/\text{GeV}\sim 10^{55}V/\text{kpc}^3 $. Since $ \nu \sim 10^{9\delta -1}\left[ \left( B/\text{\textmugreek G}\right) \left( L/\text{kpc}\right) \right] ^{\delta} $, $ T\sim \left( U-\mu N \right) /\nu $ should typically be an extremely macroscopic energy. This is consistent with our definition of subsystems, and suggests that the cutoff energy of the observational spectrum can not be ascribed to $ T $. Other mechanisms that can produce a spectral high-energy cutoff, e.g., energy losses of particles, should not change the equipartition condition, as each lost subsystem also carries the same mean energy $ T $.

The following representation is a heuristic generalization of thermal equilibrium for particles interacting with a continuous background. Let there be a number of structures continuously distributed in the system. Such a distribution can generally be expressed with a Fourier expansion, in which each Fourier component represents a set of characteristic structures with the length scale $ 2\pi /k $. For a $ \delta $-dimensional distribution in a finite volume with the length scale $ L $, the $ k $-space DOS $ d\nu /dk $ may be given by Eq.~(\ref{DOS}). If each structure can be associated with an unrestricted number of particles at a characteristic energy $ \varepsilon _{\text{c}} $, considering the overall system as a microcanonical ensemble (see the counting argument at the end of Sect.~\ref{SD}), under thermal fluctuations the mean particle number $ \overline{n} $ associated with a Fourier component should satisfy the BE distribution Eq.~(\ref{BE}) (with $ \varepsilon _{\parallel}\rightarrow \varepsilon _{\text{c}} $). The particle spectrum with respect to $ \varepsilon _{\text{c}} $ is thus $ \overline{n}d\nu /dk\left| \partial k/\partial \varepsilon _{\text{c}}\right| $, which reduces to a break PL spectrum if $ k\propto \varepsilon _{\text{c}}^{-\beta} $ (with $ \beta $ independent of $ \varepsilon _{\text{c}} $) and $ \varepsilon _{\text{c}}-\mu \ll T $. The spectral index is $ \alpha =\beta \delta +1 $ for $ \varepsilon _{\text{c}}\ll -\mu \ll T $, and $ \alpha +1 $ for $ -\mu \ll \varepsilon _{\text{c}}\ll T $. In the PA scattering problem, the structures are interpreted as the scattering waves, $ \varepsilon _{\text{c}} $ is the parallel energy $ \varepsilon _{\parallel} $, and $ \beta =1/2 $ and 1 for NR and UR particles, respectively. It is also possible to interpret the structures as particle sources, each of which can generate a particle distribution sharply peaking or cutting off at $ \varepsilon _{\text{c}} $. For small-scale sources with $ 2\pi /k\ll L $, one may replace $ \varepsilon _{\text{c}} $ directly with the particle energy $ \varepsilon $, i.e., the spectrum with respect to $ \varepsilon $ is a series of the peak or cutoff distributions. The high-energy cutoff of the $ \varepsilon $-spectrum may theoretically be at the temperature $ T $ or maximum $ \varepsilon _{\text{c}} $, or due to energy losses of particles. This representation allows us to incorporate particle acceleration processes, e.g., shock acceleration and magnetic reconnection, into the scenario of thermal equilibrium. Sources of such processes are thought to be plasma turbulence, solar particle events, supernova remnants, pulsar wind nebulae, active galactic nuclei, etc. An apparent relation between the source size and maximum energy of particles accelerated by the source can be obtained with the Hillas criterion, which claims that particles with gyro-radii greater than the source size can not be accelerated \cite{1984ARA&A..22..425H}. This leads to a $ k $-$ \varepsilon_{\text{c}} $ relation formally identical to the GR condition Eq.~(\ref{GR}), but here the magnetic field $ B $ is observed in the source region and is thus generally a function of $ k $. For $ B\propto k^{\gamma} $, one has $ \beta =1/\left[ 2\left( 1-\gamma \right) \right] $ and $ 1/\left( 1-\gamma \right) $ for NR and UR particles, respectively. Such a scenario then expects $ \gamma <1 $ for the observed average distribution of cosmic energetic particles ($ \alpha >1 $).
\section*{Acknowledgment}
I thank Drs.~Siming Liu, Dejin Wu, Heyu Sun and Xiaowei Zhou for their help. This work is partially supported by grants from the National Natural Science Foundation of China (Grant No.~11761131007, U1738122, U1931204), National Key R\&D Program of China (2018YFA0404203), International Partnership Program of Chinese Academy of Sciences (114332KYSB20170008), and China Scholarship Council (201806340077).
\bibliographystyle{unsrt}

\end{document}